\documentclass[conference]{IEEEtran}
\IEEEoverridecommandlockouts
\usepackage{algorithm}
\usepackage{algorithmic}
\usepackage[mode=buildnew]{standalone}
\usepackage{amsmath,mathtools,nccmath}
\usepackage{siunitx}
\usepackage{cite}

\usepackage{dblfloatfix}
\usepackage{booktabs}
\usepackage{multirow}

\usepackage{enumitem}
\usepackage{mathtools}
\usepackage[utf8]{inputenc}
\usepackage{amssymb}
\usepackage{gensymb}

\usepackage[compact]{titlesec}
    \titlespacing{\section}{0pt}{2ex}{1ex}
    \titlespacing{\subsection}{0pt}{1ex}{0ex}
    \titlespacing{\subsubsection}{0pt}{0.5ex}{0ex}
\usepackage{float}
\usepackage{array}
\usepackage{algorithmic}
\usepackage{textcomp}
\usepackage{xcolor}
\usepackage{tabularx}
\def\BibTeX{{\rm B\kern-.05em{\sc i\kern-.025em b}\kern-.08em
    T\kern-.1667em\lower.7ex\hbox{E}\kern-.125emX}}

\begin{document}

\include{bibliography}

\title{Hybrid Machine Learning Approach for Cyberattack Mitigation of Parallel Converters in a DC Microgrid\\
}

\author{
\IEEEauthorblockN{Naser Souri, \emph{Graduate Student Member, IEEE}, Ali Mehrizi-Sani\thanks{This work is supported in part by the National Science Foundation (NSF) under award ECCS-1953213, in part by the State of Virginia’s Commonwealth Cyber Initiative (www.cyberinitiative.org), in part by the U.S. Department of Energy’s Office of Energy Efficiency and Renewable Energy (EERE) under the Solar Energy Technologies Office Award Number 38637 (UNIFI Consortium led by NREL), and in part by Manitoba Hydro International. The views expressed herein do not necessarily represent the views of the U.S. Department of Energy or the United States Government.}, \emph{Senior Member, IEEE}}
\IEEEauthorblockA{The Bradley Department of Electrical and Computer Engineering 
Virginia Tech, Blacksburg, VA 24061 \\ \thanks{“© 20XX IEEE.  Personal use of this material is permitted.  Permission from IEEE must be obtained for all other uses, in any current or future media, including reprinting/republishing this material for advertising or promotional purposes, creating new collective works, for resale or redistribution to servers or lists, or reuse of any copyrighted component of this work in other works.”}
%\IEEEauthorblockA{$^2$Department of Energy and Control, University of Normandy-ESIGELEC-IRSEEM, Rouen, France}
e-mails: \{nsouri,mehrizi\}@vt.edu} %, tehrani@esigelec.fr}

}

\maketitle

\begin{abstract}
Cyberattack susceptibilities are introduced as the communication requirement increases with the incorporation of more renewable energy sources into DC microgrids. Parallel DC-DC converters are utilized to provide high current and supply the load. Nevertheless, these systems are susceptible to cyberattacks that have the potential to disrupt operations and jeopardize stability. Voltage instability may result from the manipulation of communication commands and low-layer control signals. Therefore, in this paper, a cyberattack that specifically targets parallel DC-DC converters is examined in a DC microgrid. A hybrid machine learning-based detection and mitigation strategy is suggested as a means to counteract this threat. The false data injection (FDI) attack targeting the converters is investigated within a DC microgrid. The efficacy of the suggested approach is verified via simulations executed for various scenarios within the MATLAB/Simulink environment. The technique successfully identifies and blocks FDI attacks, preventing cyberattacks and ensuring the safe operation of the DC microgrid. 
\end{abstract}

% Cyberattack susceptibilities are introduced as the incorporation of renewable energy sources into DC microgrids expands. This paper investigates a cyberattack targeting parallel DC-DC converters, which are utilized to supply and regulate power flow within the microgrid. However, these systems are vulnerable to cyberattacks that can disrupt operations and potentially compromise stability. The attack manipulates the control signals of the converters and communication commands, potentially leading to voltage instability. To address this threat, a dual machine learning detection and mitigation method is proposed. False Data Injection (FDI)  attack targeting the control loops of parallel converters within a DC microgrid is investigated. The effectiveness of the proposed method is validated through simulations conducted in a MATLAB/Simulink environment. The simulation results demonstrate the method ability to successfully detect and mitigate the FDI attack, ensuring stable operation of the DC microgrid under cyber threats.


%=======================================================
\section{Introduction}
\begin{figure}[!t]
\centerline{\includegraphics[width= 0.6\columnwidth ]{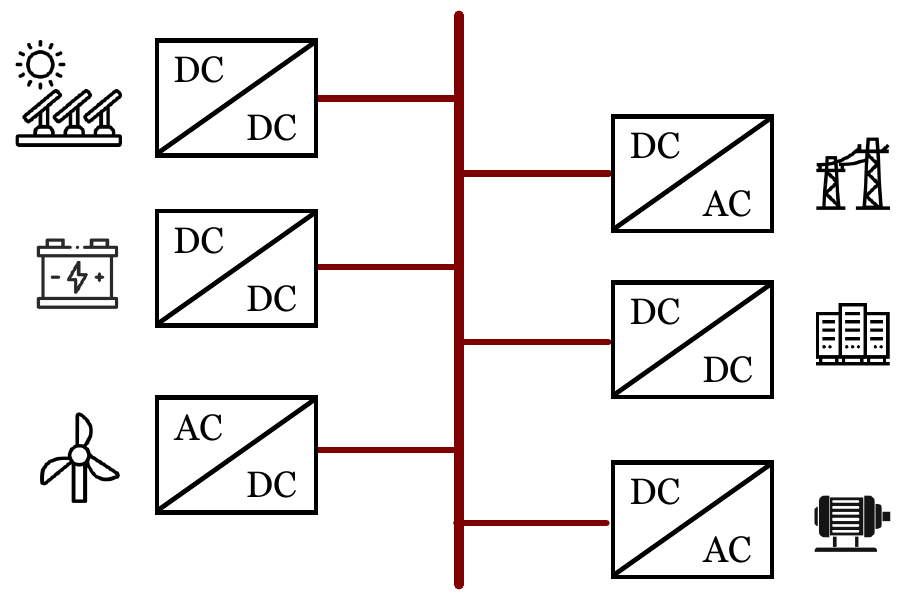}}
\vspace*{-0.2 cm}
\caption{General block diagram of a DC microgrid topology.}
\label{fig:DC_microgrid1}
\end{figure}

DC microgrids are emerged as a promising solution for integrating distributed renewable energy sources and energy storage systems into the power grid.
%In addition to reducing the reliance on fossil fuels, this method promotes the use of renewable energy sources. 
\cite{Farhangi}. Renewable sources generate DC power directly; therefore, they eliminate the need for conversion losses associated with AC transmission. Nevertheless, the extensive implementation of DC microgrids encounters several obstacles. An important obstacle is the guarantee of secure operation and control of the microgrid \cite{Harshbarger}. A general block diagram of a DC microgrid is shown in Fig.~\ref{fig:DC_microgrid1}.

%The reliance on communication networks and control systems introduces new vulnerabilities to cyberattacks.
%Ensuring the safe operation of a DC grid presents a unique challenge in the grid. 

As the demand for electricity increases and additional distributed sources are deployed, parallel converters become essential for effectively managing this evolving grid,  which offers several benefits, such as modularity and scalability, and promotes system redundancy in the grid.
If one converter malfunctions, the others can continue to operate, maintain operation, and minimize interruptions. 
%This fault tolerance improves the overall reliability and availability of the microgrid. 
However, the increased complexity of a system with multiple interconnected converters requires resilient control and communication mechanisms in a smart grid. %A smooth grid distribution relies on remote communication between utilities and DERs for control 
%Unfortunately, cyberattacks compromising this communication can cripple a utility ability to manage the grid effectively.
% Despite their numerous advantages, DC microgrids present a new frontier for cyberattacks. 
Reliance on communication networks and control systems for data exchange renders them susceptible to cyber threats \cite{Chengcheng}. By compromising communication networks or sensors, it could alter control signals and system efficiency or cause voltage instability and potentially damage the equipment \cite{Gholami}. Attacks on sensors or communication can have catastrophic consequences in cyber-physical systems (CPS). A successful DC microgrid hack might cause localized disturbances or widespread blackouts. Therefore,  DC microgrid integrity depends on strong cybersecurity, such as encryption and intrusion detection. To tackle these cyber threats, many researchers examine strategies for the detection and mitigation of cyberattacks. Cyberattacks targeting key power grid parameters can lead to system deficiencies and economic losses \cite{Beikbabaei}

Smart grids face a diverse range of cyber threats. Denial-of-service (DoS), false data injection attack (FDIA), man-in-the-middle (MITM), malware injection, delay attacks, and signal jamming all are various types of cyberattacks \cite{Bhusal, Teymouri}. FDIA and DoS attacks are the most common types of attack. Attackers might launch FDIA on a cyber-physical system by modifying remote terminal unit (RTU) firmware \cite{TAHER ,Milad }. 

Model-based or data-driven methods are two traditional ways to tackle these challenges, especially FDIA \cite{Jhala}. Model-based approaches discover system abnormalities using the system model generated through state estimates. For example, \cite{Madichetty} proposes a virtual sensor-based structure for malicious activities and designs an adaptive state observer for voltage estimation in DC microgrids. Reference \cite{Tariq} proposed a detection method using the Kalman filter observer-based in DC microgrids. However, it only considers the attack on the communication link between the sources. The observer-based approach necessitates a precise representation of the entire system model, which may not be feasible practically in all cases. In addition, a stealth attack on CPS could bypass the conventional observer-based detection method \cite{Sahoo}.
%Various methods are employed to detect and mitigate cyber threats including model-based and data-driven.

\begin{figure}[!t]
\centerline{\includegraphics[width= 0.7\columnwidth ]{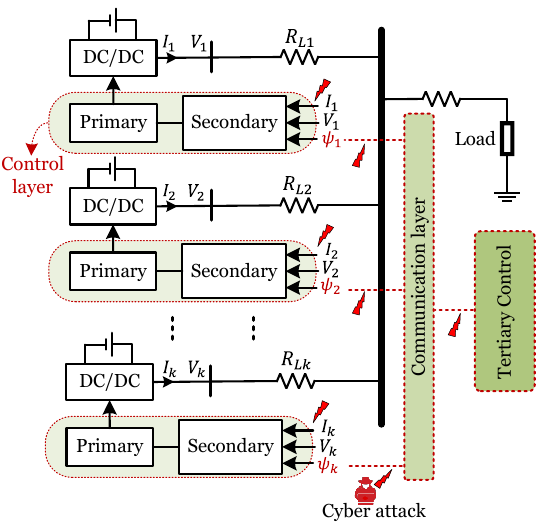}}

\vspace*{-0.2 cm}
\caption{Architecture of parallel DC-DC converters with communication lines.}
\label{fig:Architecture}

\vspace*{-0.3 cm}
\end{figure}

On the other hand, the data-driven technique detects cyberattacks using the device's voltage and current measurements. In recent years, data-driven machine-learning approaches have been used to detect anomalies and estimate parameters to replace fraudulent data. Classification techniques such as K-nearest neighbor (KNN), which are suggested in \cite{Atefi, Hasnat}, can be used for FDIA detection. Artificial neural networks (ANN) outperform KNN for anomaly detection and can be used for more complex malicious signals instead of KNN. ANN can be used for both estimation and classification, where there are many unknown input data points. For example, reference \cite{Blaabjerg} uses ANN combined with model predictive control to detect anomalies while there are cyber threats. Reference \cite{Sahoo} introduces criteria for each agent to address stealth attacks, with an emphasis on the regulation of the average voltage in a DC microgrid. Reference \cite{Shuai} attempts to detect FDIA using artificial intelligence (AI). This technique detects FDIA on voltage and current sensors using an ANN technique.

Long-short-term memory (LSTM) networks are a type of ANN architecture that excels in forecasting and anomaly detection. They are ideal for evaluating time-series data. By analyzing sequences of network events, LSTM can learn network behavior and estimate the parameters. On the other hand, hybrid ML-based methods, which are becoming more attractive, can increase detection accuracy and reduce errors during load changes or in noisy systems. Some of the hybrid methods are reported in \cite{Sharma, Presekal, Popoola}. Accuracy comparisons for individual and hybrid ML-based models are provided in \cite{Sharma}. 

Therefore, this paper uses a hybrid ML-based method to detect cyberattacks, estimate falsified data, and replace it with estimated values. The contribution of this paper is as follows:
\begin{itemize}
  \item Proposing a hybrid ML-based algorithm for detection and mitigation of FDI attacks on communication lines and sensors for parallel converters.
\end{itemize}

\begin{figure}[!t]
\centerline{\includegraphics[width= 0.6\columnwidth ]{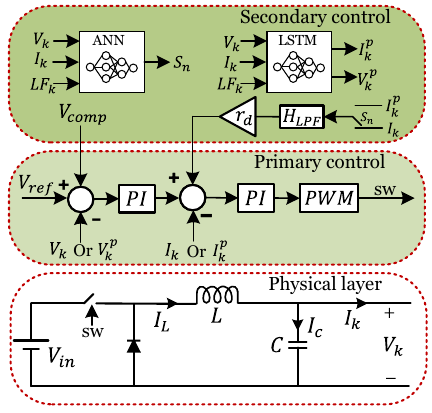}}
\vspace*{-0.2 cm}
\caption{Physical and control layers of a DC-DC converter.} 
\label{fig:control}

\vspace*{-0.2 cm}
\end{figure}

\begin{figure}[!t]
\centerline{\includegraphics[width= 0.5\columnwidth ]{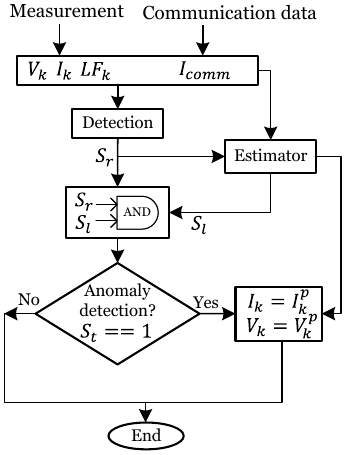}}
\vspace*{-0.2 cm}
\caption{Detection and mitigation algorithm.}
\label{fig:algorithm}

\vspace*{-0.2 cm}
\end{figure}

The rest of the papers are as follows.
Section II provides the system description under study. Section III delves into the proposed method for detection and mitigation based on hybrid machine learning. Section IV provides simulation studies in various scenarios to evaluate the performance of the proposed method. Section V concludes the hybrid machine learning method.

%=======================================================
\section{System Description}
Fig.~\ref{fig:Architecture} shows the system architecture for this study. $k$ DC sources are considered in parallel, each of which is connected to the DC bus using a DC-DC converter with equal rating power. The microgrid parameters are shown in Table I. $V_k$, and $I_k$ represent the output current and voltage of each converter, and $V^p_k$ and $I^p_k$ represent the estimated values for the current and voltage, respectively. The line resistors ($R_L$) are considered different as, in practice, each source may be located at different distances. A hierarchical structure is used for the converter control and operation. Each converter is equipped with a cascade control in the primary layer to control both voltage and current. The secondary layer is responsible for bus voltage regulation, power sharing, anomaly detection, and mitigation, which are discussed in the next sections. Fig.~\ref{fig:control} shows the DC-DC topology, the primary control, and the secondary layer in this study. The primary control consists of voltage and current control. Power sharing is achieved using the average current of the converters via communication lines. $V_{ref}$ is the reference voltage for the converters. $r_d$ is the droop gains for equal power sharing. $H_{LPF}$ is a low-pass filter to smooth the output current for power sharing purposes. 

The transmitted data comes from neighbors, and measured values for the primary control come directly from the sensors, or predicted values may be replaced if an attack is launched. The output current of the converters is transmitted using the communication link between the converters, which is considered a distributed structure. Unlike centralized architecture, distributed architecture avoids a single point of failure. Furthermore, using distributed communication, the other converters can be informed about the manipulated data to prevent instability.

\begin{table}[t]
\caption{System Parameters}

\begin{center}
\begin{tabular}{lcclcc}

\toprule
\textbf{Parameter} & \textbf{Symbol} & \textbf{Value}  & \textbf{Parameter}  & \textbf{Symbol} & \textbf{Value} \\ 
\toprule
\toprule

\begin{tabular}[c]{@{}l@{}}Input \\ voltage\end{tabular}         & $V_{in}$              & \begin{tabular}[c]{@{}l@{}}80 V\\ \end{tabular} & \begin{tabular}[c]{@{}l@{}}Output\\ voltage \end{tabular}    & $V_O$              & 39 V         \\ 
%\midrule
%\begin{tabular}[c]{@{}l@{}}Inductor \\ resistor\end{tabular}      & $R_{L}$             & 50 $m\Omega$                                                 & \begin{tabular}[c]{@{}l@{}}Capacitor\\ resistor\end{tabular} & $R_{c}$              & 5 $m\Omega$         \\ 
\midrule
\begin{tabular}[c]{@{}l@{}}Rated \\ power \end{tabular} & $P_{out}$              & 150 W                                                 & \begin{tabular}[c]{@{}l@{}} Inductor \\ \end{tabular}        & $L_{buck}$              &  2 mH         \\ 
\midrule
\begin{tabular}[c]{@{}l@{}}Capacitor\\  \end{tabular}   & $C_{buck}$             & 100 uF                                                & \begin{tabular}[c]{@{}l@{}}Switching\\ frequency\end{tabular}   & $F_s$              & 25 kHz        \\ 
\midrule
\begin{tabular}[c]{@{}l@{}}Line 1\\resistor  \end{tabular}   & $R_{L1}$             & 0.7 $\Omega$                                                & \begin{tabular}[c]{@{}l@{}}Line 2\\resistor\end{tabular}   & $R_{L2}$              & 0.6 $\Omega$        \\ 
\midrule
\begin{tabular}[c]{@{}l@{}}Line 3\\resistor \end{tabular}   & $R_{L3}$             & 0.5 $\Omega$                                                & \begin{tabular}[c]{@{}l@{}}Line 4\\resistor\end{tabular}   & $R_{L4}$              & 0.4 $\Omega$        \\ 
\bottomrule
\end{tabular}
\label{table:system_parameters}
\end{center}

\vspace*{-0.2 cm}
\end{table}

\section{Detection and Mitigation Technique}
In this section, the proposed hybrid machine learning method is discussed. 
%As mentioned in the literature review, the detection and mitigation of a cyber attack using only one machine learning cell is less accurate in practice for both purposes and is not able to detect all cases, e.g., it may have an error in detection in load change or a noisy system using only one machine learning. In this paper, therefore, dual ML-based method are utilized for this purpose to reduce detection errors in the system. 
Fig.~\ref{fig:algorithm} shows the proposed algorithm, including the detection and estimation blocks. Logistic regression is used for falsified data detection, and long-short-term memory networks are used for the prediction of falsified data and detection as well. These two blocks are trained separately to increase the accuracy of detection and estimation. If there is an error in the detection, then using estimated data may introduce more oscillation in the system. However, these errors are reduced using the hybrid method.

\begin{figure}[!t]
\centerline{\includegraphics[width= 0.65\columnwidth ]{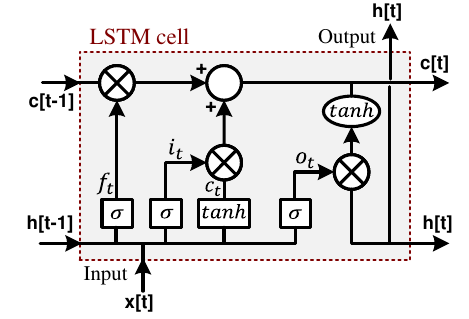}}

\vspace*{-0.3 cm}
\caption{The LSTM cell block diagram.}
\label{fig:LSTM}

\vspace*{-0.3 cm}
\end{figure}

According to Fig.~\ref{fig:algorithm}, the detection block gets the measured values from the local converter and from the communication, and based on these data and the assigned binary flags for all parameters, it decides which parameter is false (anomaly detection) or true (normal condition). Therefore, the output state of the detection block is a binary vector $S_r$, where $S_r = 0$ means no detection and $S_r = 1$ means it detects an anomaly. This vector includes all measurements and transmitted data, for example, $S_r=[S_c, S_v, S_{com-1}, S_{com-2}]$, such that $S_c$, $S_v$, and $S_{com-i}$ represent anomaly detection flags for current, voltage, and communication data from the $i^{th}$ converter in the detection block, respectively. One of the key advantages of this approach is that the parameters that are attacked are recognizable and can be replaced with their estimated value. 

After anomaly detection, the LSTM cell generates another binary vector, $S_l$, by comparing the estimated value with the falsified data that is detected by regression for anomaly detection confirmation. $S_l$ vector is the same as $S_r$ (e.g., $S_l=[S'_c, S'_v, S'_{com-1}, S'_{com-2}]$, such that $S'_c$ and $S'_v$, $S'_{com-i}$ represent anomaly detection flags for current, voltage, and communication data from the $i^{th}$ converter in the LSTM block. If both $S_r$ and $S_r$ agree on a detection command, then the estimated data is replaced by the measured or transmitted data. $S_t$ changes from zero to one if both $S_l$ and $S_r$ are one. For example, if $S_c$ and $S'_c$ change their value from zero to one, it means that an attack occurs only on the current.

In the training process, it is crucial to separate the data into training and test sets for each machine learning block. The number of hidden layers for the LSTM is four, and 20,000 points of the measured data are sampled for machine learning training, among which 70\% of the sampled data is used for training, 15\% is used for data validation, and 15\% is used for testing the data to avoid overfitting. The data is trained within 234 epochs using the Levenberg-Marquardt algorithm, and the training performance is evaluated with mean square error (MSE), which reaches less than 2.016e-4.
In the following subsection, the machine learning tools are discussed.

%After detection from both blocks, the LSTM cell replaces only the falsified parameters instead of all of them. 
%Therefore, hybrid ML ensures the safe operation of the microgrid. 

% $S_j$ is an input for the LSTM estimator along with the other data. %A simple two-layer ANN is used for detection and Long Short-Term Memory networks are used for estimation. 
\subsection{Regression and LSTM Tools}
Logistic regression is a fast machine learning algorithm that can be used for classification tasks, particularly binary classification. Thus, this approach allows for the identification of normal or abnormal conditions in cyberattack scenarios \cite{Ray}. Over 96\% accuracy is achieved using the regression for anomaly detection reported in \cite{Bhusal}. Thus, it is utilized as a detection tool in the algorithm shown in Fig.~\ref{fig:algorithm}. On the other hand, LSTM is a type of ANN architecture specifically designed to excel at handling time series data, which can learn long-term dependencies in sequential information \cite{Beikbabaei}. Over 96.5\% accuracy is reported in \cite{Khan} for anomaly detection using the LSTM tool. The LSTM estimator block diagram is shown in Fig.~\ref{fig:LSTM}. Three gates comprise the LSTM cell: input, output, and forget gates. Information is classified as critical or unimportant by these gates. $f_t$ is the forget gate, which can be either zero or one and is determined by:

\begin{equation}
f_t = \sigma(w_f[h_{t-1},x_t] + b_f).
\end{equation}

The input and output gates are represented as:

\begin{equation}
\begin{split}
& i_t = \sigma(w_i[h_{t-1},x_t] + b_i,  \\
& o_t = \sigma(w_o[h_{t-1},x_t] + b_o.  
\end{split}
\end{equation}

\noindent where $w_f$, $w_i$ and $w_o$ are the weights for the gate neurons, and $b_f$, $b_i$ and $b_o$ are the biases for the gates. The input gate and forget gate update the cell state $c_t$. $h[t-1]$ and $c[t-1]$ represent the previous step of their values. The output $h[t]$ is updated using the cell state and output gate. $tanh$ is a hyperbolic tangent function and $\sigma$ is the Sigmoid function:

\begin{equation}
\sigma = \frac{1}{1 + e^{-x}}.
\end{equation}

% Machine learning detection and estimation are trained under various attacks and normal conditions. 20,000 data are sampled for the train. 

% The accuracy of the detection using regression and LSTM is over 96.69\% and 96.78\%, respectively \cite{Bhusal}, and the accuracy of the hybrid method increases to over 99\%. 
In the next section, the proposed machine learning-based algorithm for cyberattack detection is validated using simulation studies.

% ================================================

\section{Simulation Studies}
To validate the proposed hybrid machine learning method for the detection of FDI against sensors and communication lines, simulation studies are performed in MATLAB/Simulink. A DC microgrid test bed consisting of four parallel DC-DC converters is created, similar to Fig.~\ref{fig:Architecture} with the specifications in Table I. The value of the cable resistors is considered as $R_{L1}=0.7$~$\Omega$, $R_{L2}=0.6$~$\Omega$, $R_{L3}=0.5$~$\Omega$, $R_{L4}=0.4$~$\Omega$. The simulation is performed in discrete time, and the Tustin method is used for the discretization of the PI controllers. The evaluation is performed in different scenarios below.

\begin{figure}[t]
\centerline{\includegraphics[width=1\columnwidth ]{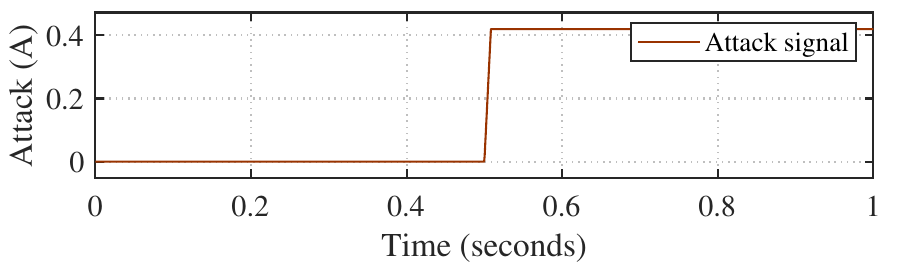}}

\vspace*{-0.4 cm}
\caption{Scenario I: the bias attack signal on the sensor for the first converter.}
\label{fig:attacksignal1}
\end{figure}

\begin{figure}[t]
\centering
\includegraphics[width=1\columnwidth]{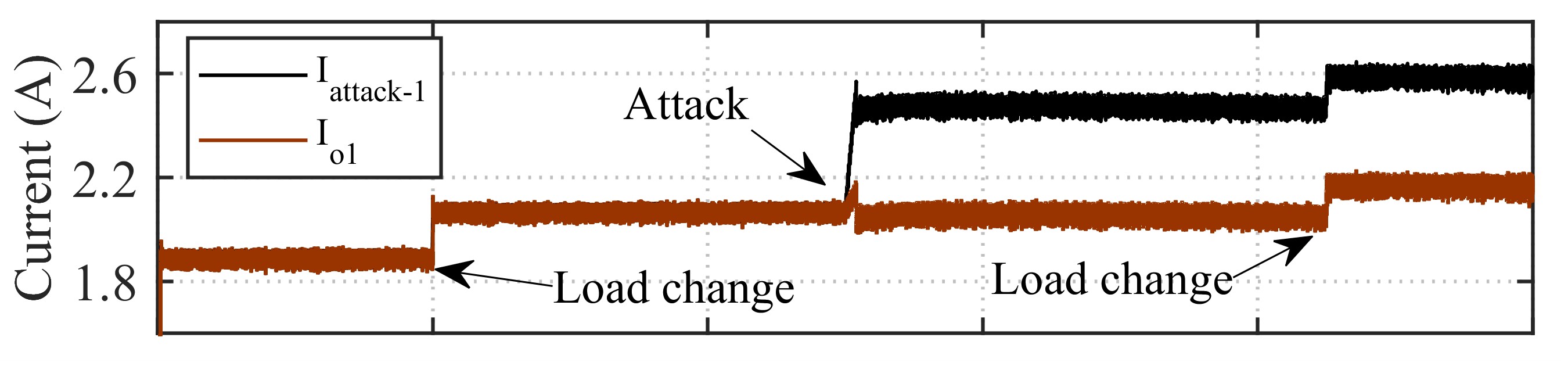}

\vspace*{-0.0 cm}
\includegraphics[width=1\columnwidth]{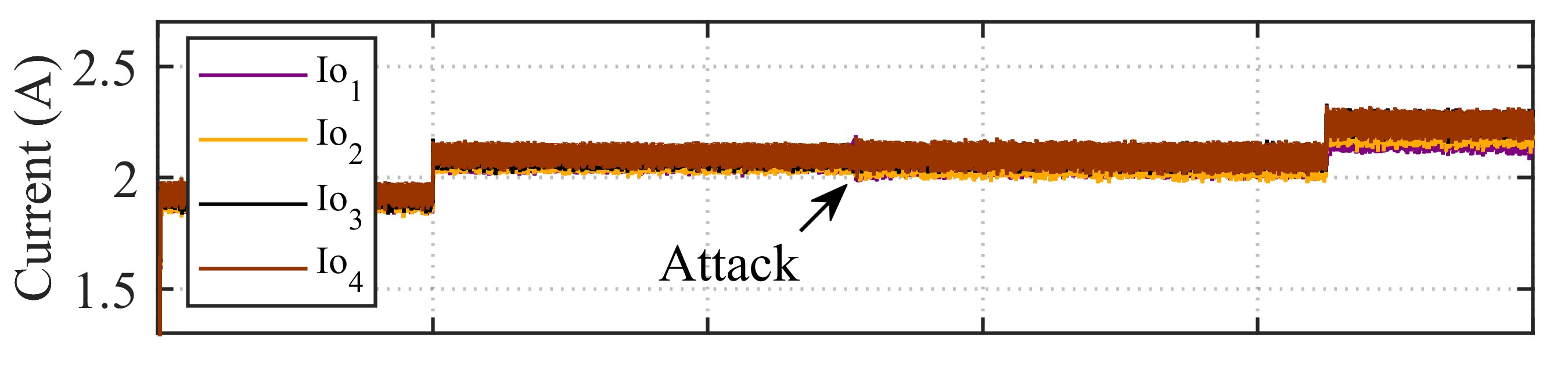}

\vspace*{-0.0 cm}
\includegraphics[width=0.98\columnwidth]{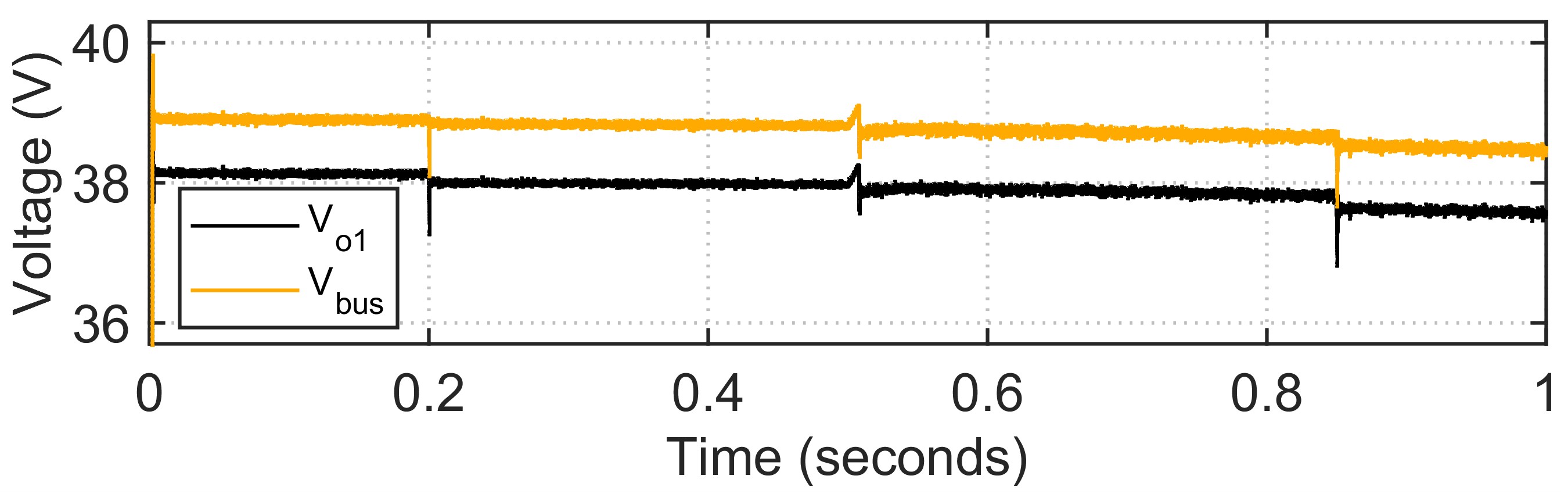}

\vspace*{-0.44 cm}
\caption{Scenario I: currents and voltages waveform of the converters and the DC bus before and after a bias attack with load change tests.}

\vspace*{-0.1 cm}
\label{fig:case1}

\vspace*{-0.3 cm}
\end{figure}

\begin{figure}
\centering
\includegraphics[width=1\columnwidth]{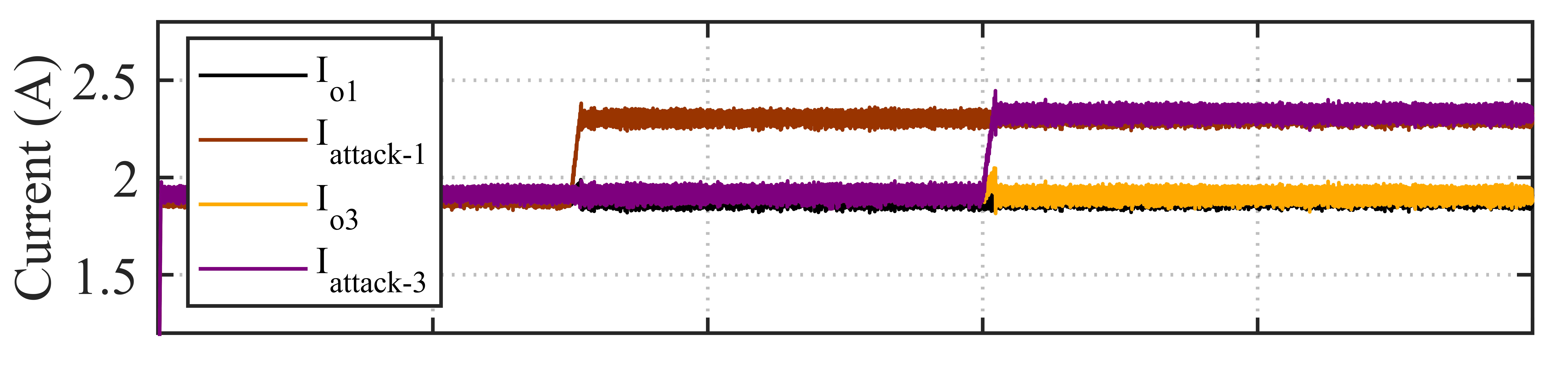}

\vspace*{-0.0 cm}
\includegraphics[width=1\columnwidth]{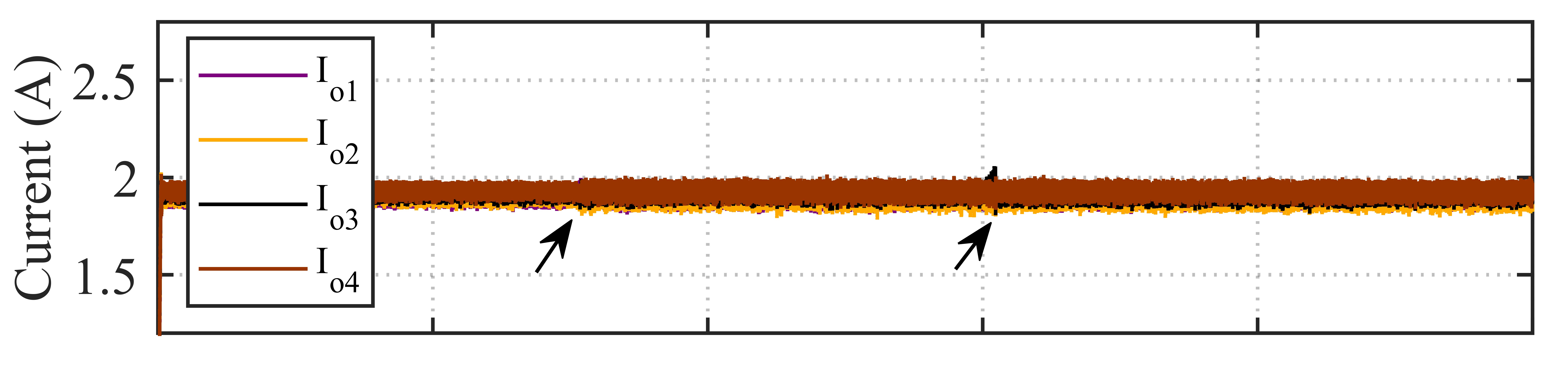}

\vspace*{-0.0 cm}
\includegraphics[width=0.98\columnwidth]{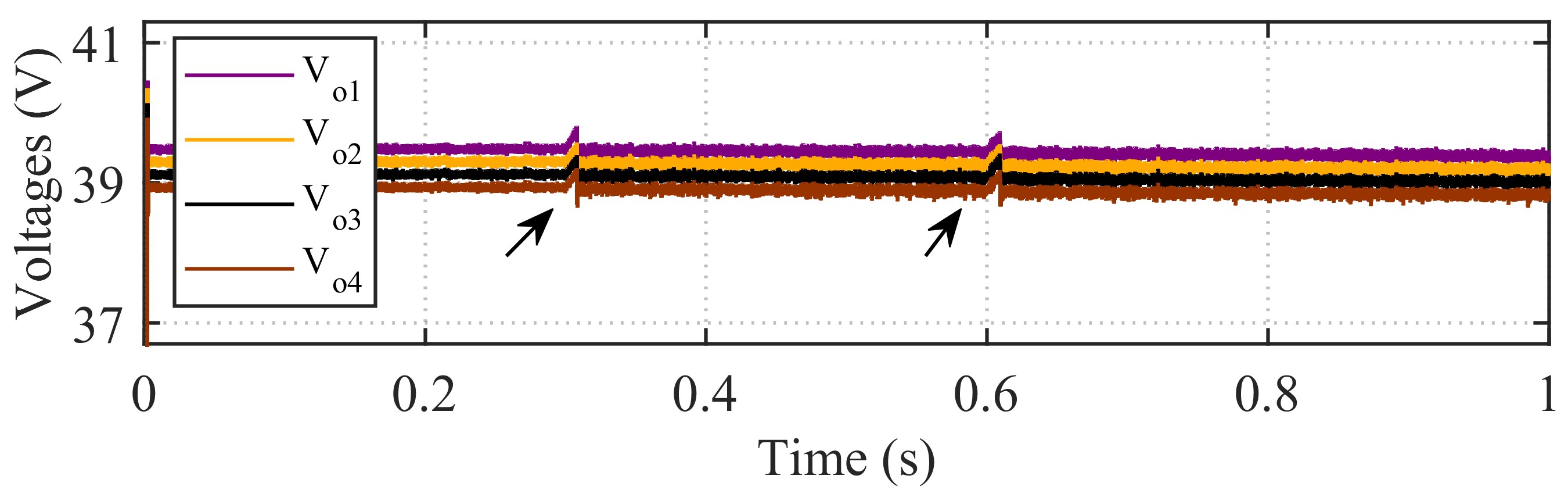}

\vspace*{-0.4 cm}
\caption{Scenario II: currents and voltages waveform of the converters and the DC bus before and after two bias attacks on two sensors.}

\vspace*{-0.1 cm}
\label{fig:case2}

\vspace*{-0.2 cm}
\end{figure}

\subsection{Scenario I: Attack on a Sensor}
In this scenario, an attack is launched on the sensor of the first converter. It is assumed that the attacker infiltrates the control layer and adds an additional value to the measurement to deteriorate the power sharing operation or damage the converter. Hence, the measurement experiences a sudden increase in the measured sensor value at $t=0.5$~s. The attack signal for this scenario is shown in Fig.~\ref{fig:attacksignal1}, which is a sharp change in the measured value of the sensor. The attack signal for this scenario is shown in Fig.~\ref{fig:attacksignal1}, which is a sharp change in the measured value of the sensor, and the results for this scenario are shown in Fig.~\ref{fig:case1}. This figure shows the converter current, with and without mitigation of the attack, all the converter output current, the converter voltage, and the bus voltage. $I_{attack-1}$ is the output current of the first converter when the attack is not mitigated, and $I_{o1}$ is the one with detection and mitigation. In this scenario, before the attack occurs, a step change in the load occurs at $t=0.2$~s to see if the proposed algorithm detects it as a normal condition or an anomaly. The results show a successful increase in the load current. To further investigate the accuracy of the algorithm, another load change is applied to the system at $t=0.85$~s while the attack occurred before and persists in the system. In both cases for load change, the algorithm recognizes it as a normal condition, which shows the effectiveness of the proposed method. Another result of this simulation is the current oscillation when the estimated value is utilized instead of the real data. As can be seen, by replacing the estimated data with the real data, the current oscillation increases. This means that if the detection block accuracy is low, the power quality degrades. Therefore, hybrid ML-based systems can increase power quality by increasing detection accuracy.

% Case 2 =============================
\subsection{Scenario II: Attack on Two Sensors}
\begin{figure}[!t]

\vspace*{-0.2 cm}
\centerline{\includegraphics[width= 1\columnwidth ]{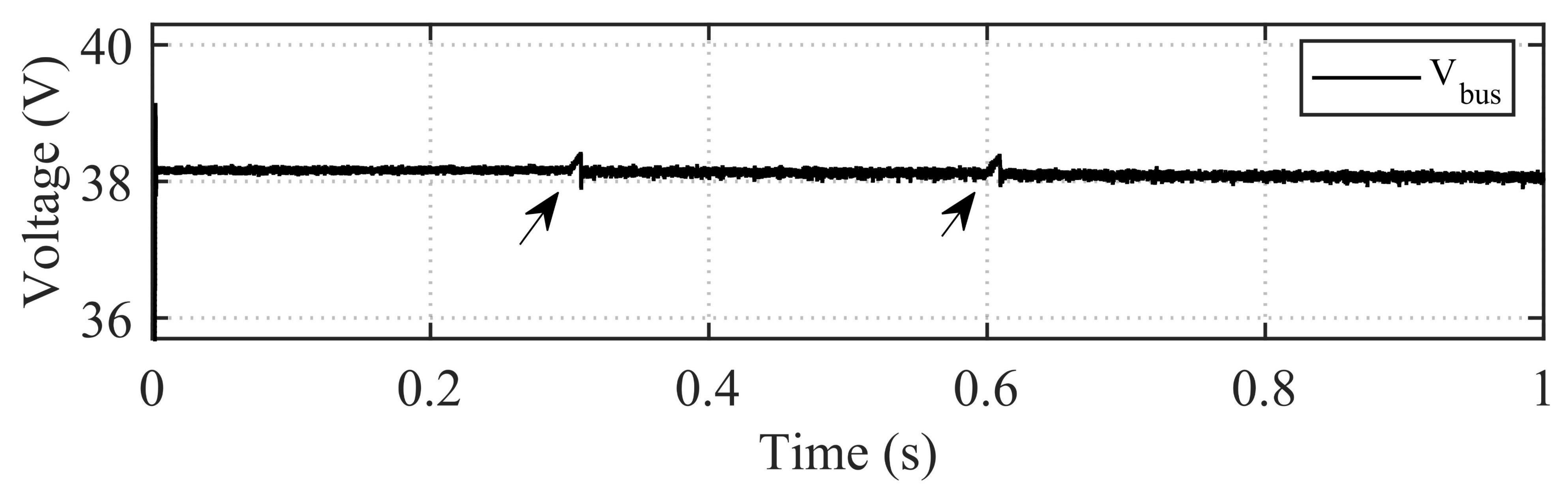}}

\vspace*{-0.4 cm}
\caption{Scenario II: Bus voltage profile after two bias attacks on two sensors.}
\label{fig:case2bus}

\vspace*{-0.1 cm}
\end{figure}

This scenario is similar to the first scenario, though two sensors from the $1^{th}$ and $3^{th}$ converters are the target of the attacker. Hence, the measurements for the first and third converters are manipulated. At $t=0.3$~s, the first sensor sends falsified data to the control layer and persists on it. Then, another sensor starts sending false data to the control layer at $t=0.0.6$~s. The results for current and voltage are shown in Fig.~\ref{fig:case2}. $I_{attack-1}$ and $I_{attack-3}$ are the output currents of the $1^{th}$ and $3^{th}$ converters, respectively, when the attack is not mitigated, and $I_{o1}$ and $I_{o3}$ are the ones with detection and mitigation. The output currents and output voltages for all converters are shown in Fig.~\ref{fig:case2}. As the results show, under such conditions, the algorithm successfully detects the FDI attack and replaces an estimated value for the manipulated data. Therefore, before a big change in the voltage and current, the attack is detected and mitigated. For this scenario, the bus voltage is shown in Fig.~\ref{fig:case2bus}. This figure shows a small change in the system, but it restores to the normal condition as the voltage deviates from a certain value of the normal condition.

% Case 4 ==========================
\subsection{Scenario III: Attack on the Communication}
\begin{figure}

\vspace*{-0.2 cm}
\centering
\includegraphics[width= 1.0\columnwidth]{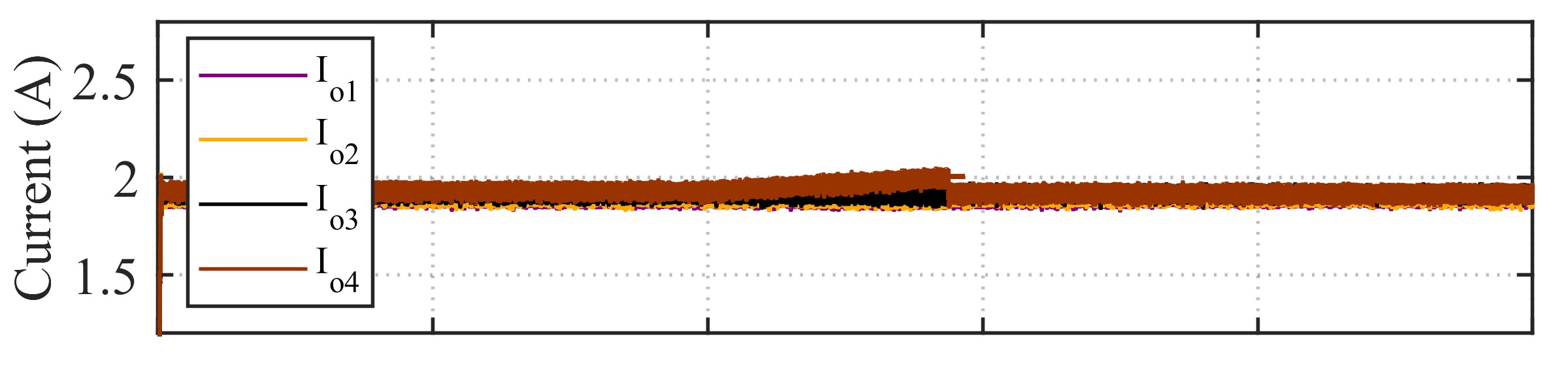}

\vspace*{-0.0 cm}
\includegraphics[width=0.98\columnwidth]{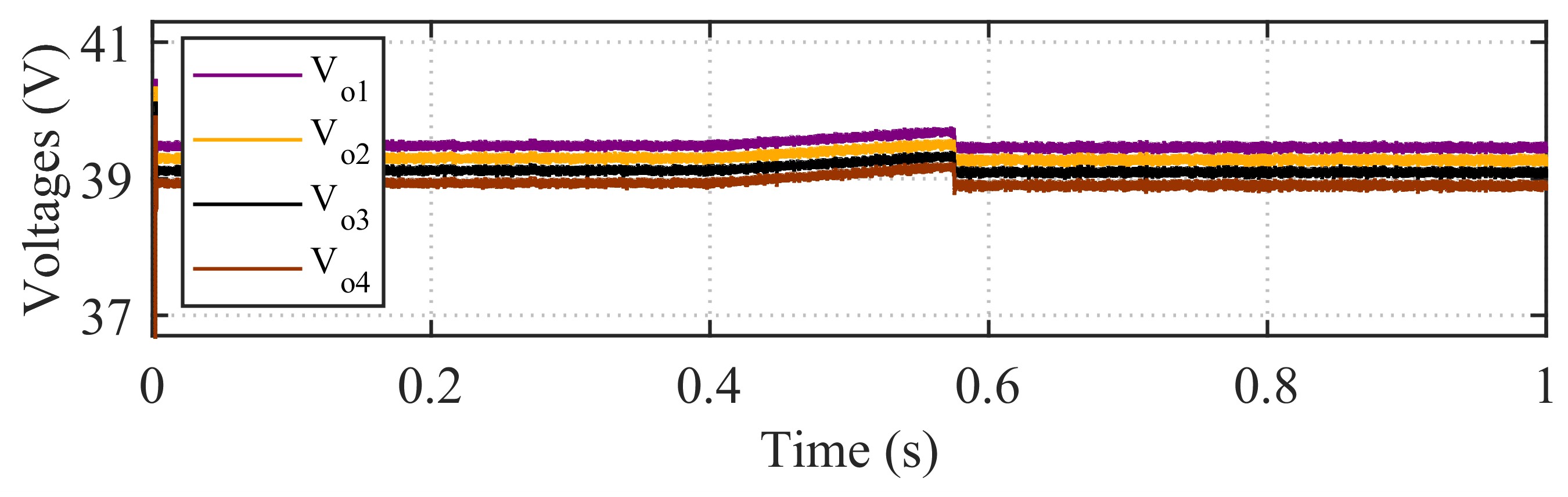}

\vspace*{-0.4 cm}
\caption{Scenario III: output currents and voltages waveform of the converters before and after a ramp attack on the communication line.}

\vspace*{-0.1 cm}
\label{fig:case4}
\end{figure}

\begin{figure}[!t]

\vspace*{-0.2 cm}
\centerline{\includegraphics[width= 1\columnwidth ]{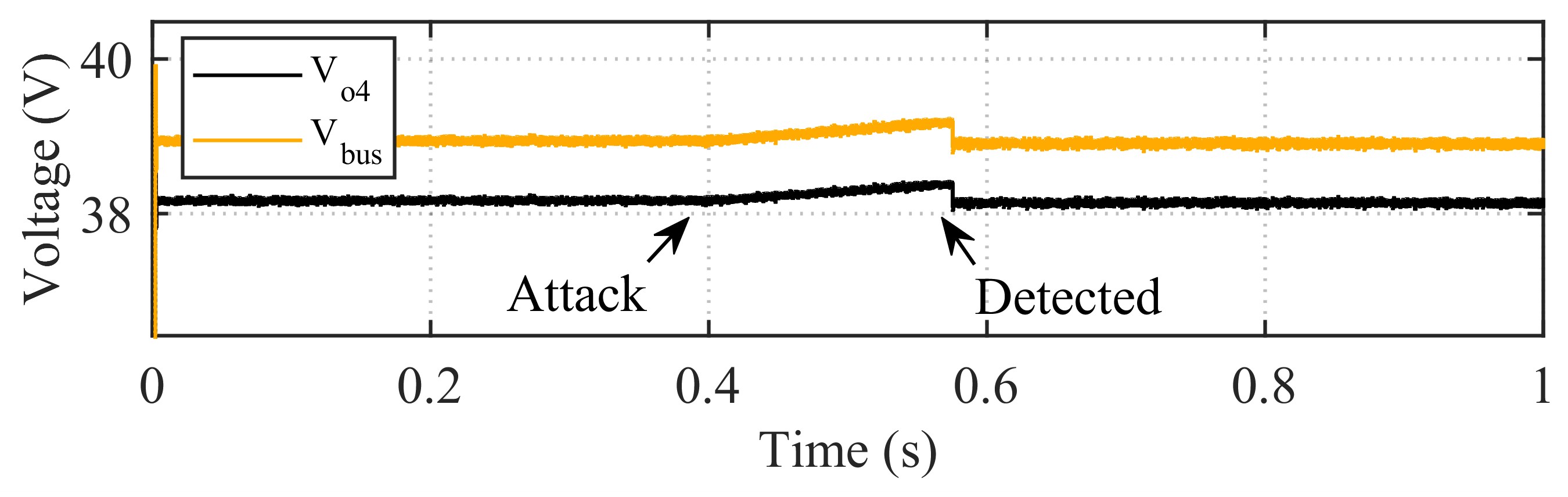}}

\vspace*{-0.4 cm}
\caption{Scenario III: bus and output voltage for the $4^{th}$ converter after a ramp attack at $t=0.4$~s.}
\label{fig:case4_bus}

\vspace*{-0.4 cm}
\end{figure}

\begin{figure}[!t]

\vspace*{+0.1 cm}
\centerline{\includegraphics[width= 1\columnwidth ]{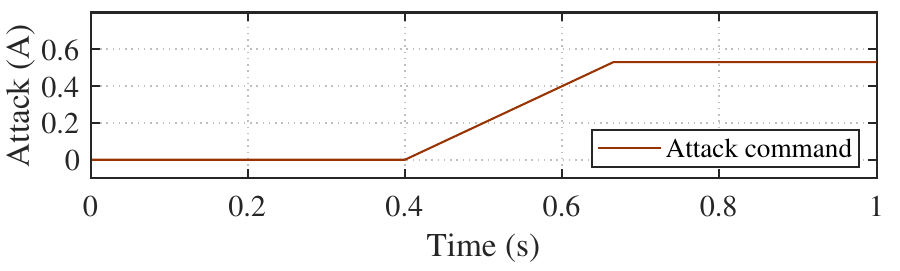}}

\vspace*{-0.4 cm}
\caption{Scenario III: the ramp attack profile on communication line.}
\label{fig:siganl4}

\vspace*{-0.3 cm}
\end{figure}

In this scenario, it is assumed that the attacker can penetrate the communication lines. In this case, a gradual change or a ramp attack is applied to the communication line. Therefore, the current for this converter starts to deviate from the normal condition. After a while, the algorithm detects the anomaly and mitigates it by replacing the false data with the estimated data. Fig.~\ref{fig:case4} shows the output currents and the voltages for all converters. The results show that by changing the transmitted data, both voltages and currents become involved and start to deviate from the normal condition; however, the algorithm detects it. Fig.~\ref{fig:siganl4} shows the attack signal profile for this scenario. Fig.~\ref{fig:case4_bus} depicted the output and bus voltage for this scenario. The voltage deviates from normal until the algorithm detects the intrusion and returns it to the normal condition.

% Case 5 ==========================
\subsection{Scenario IV: Attack on Both Sensor and Communication}

\begin{figure}[t]
\centering

\vspace*{-0.2 cm}
\includegraphics[width= 1\columnwidth ]{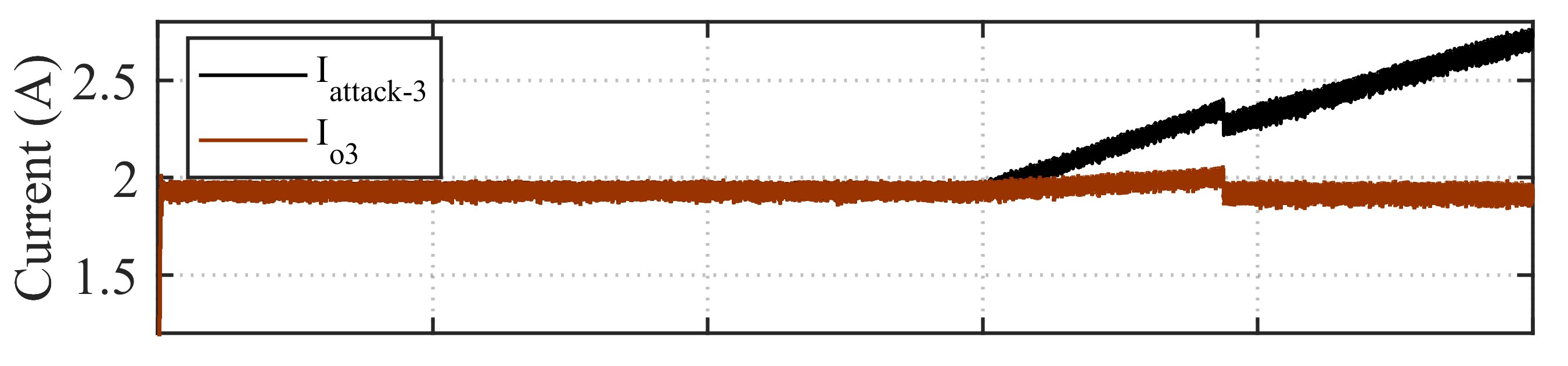}

\includegraphics[width=1\columnwidth]{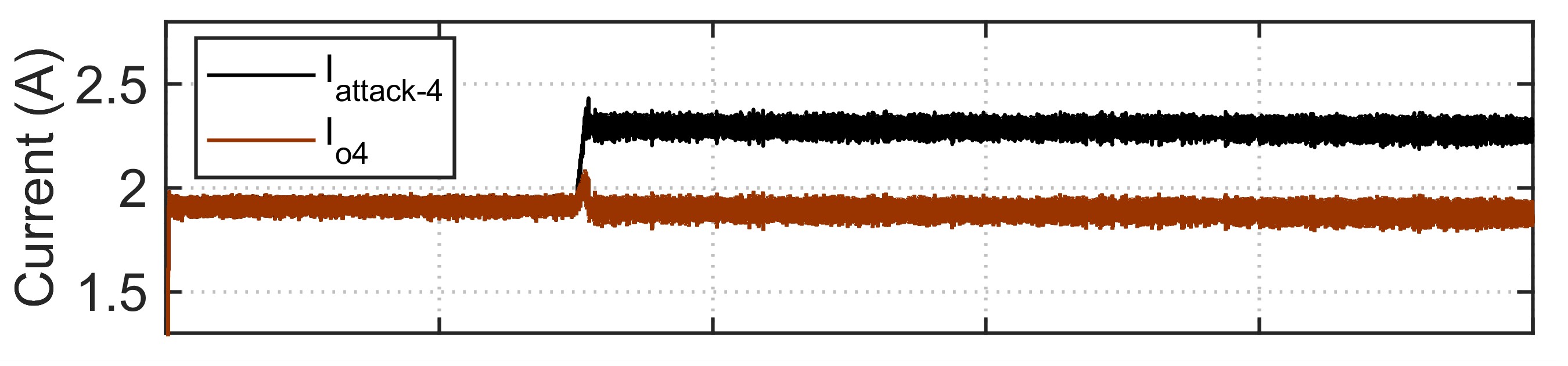}

\vspace*{-0.0 cm}
\includegraphics[width= 1\columnwidth ]{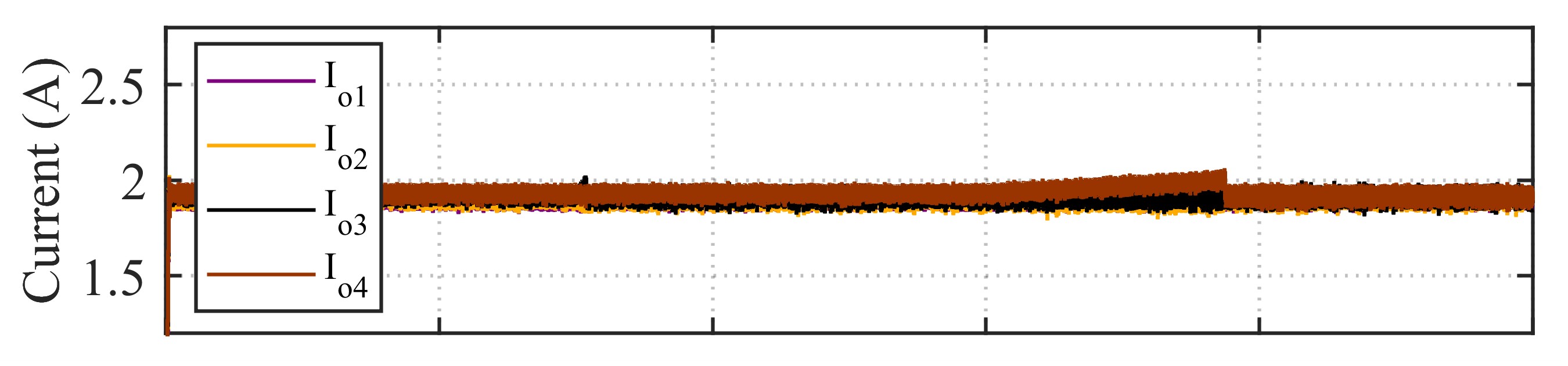}

\vspace*{+0.0 cm}
\includegraphics[width=0.98\columnwidth]{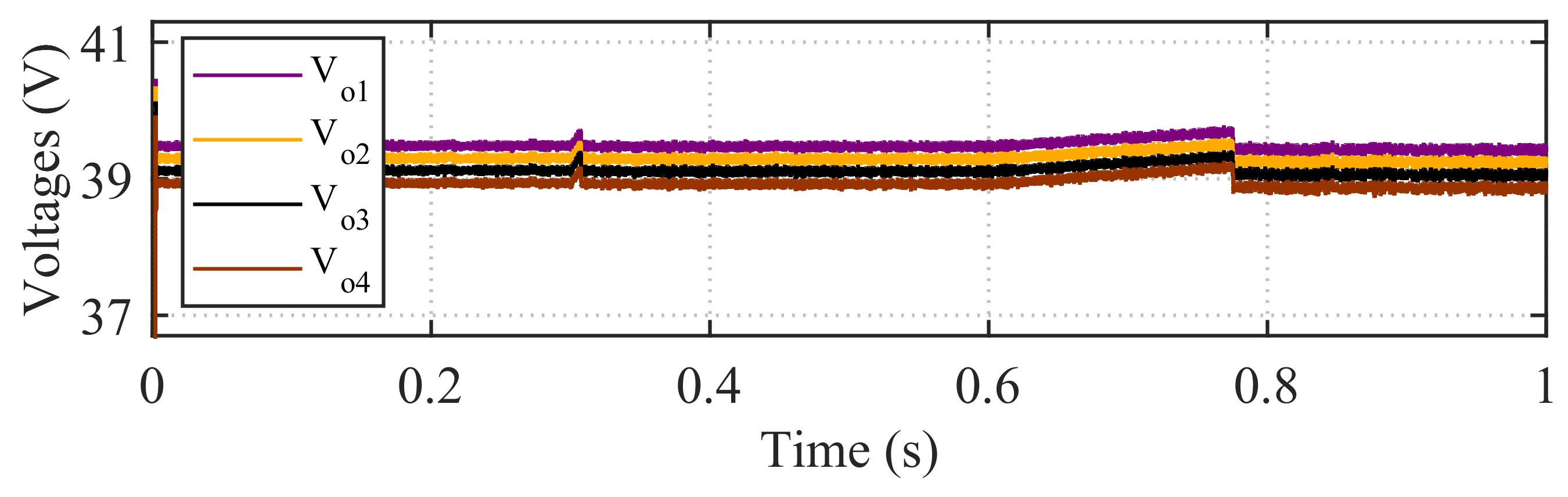}

\vspace*{-0.3 cm}
\caption{Scenario IV: currents and voltages waveform of the converters before and after a ramp attack on the communication, plus a bias attack on a sensor.}

\vspace*{-0.2 cm}
\label{fig:cordinatedramp5}
\end{figure}

\begin{figure}[!t]
\centerline{\includegraphics[width= 1\columnwidth ]{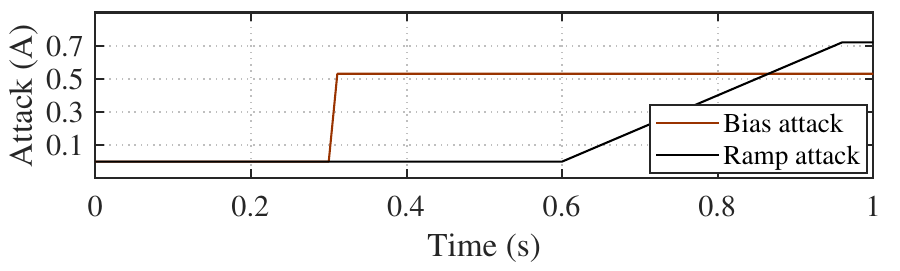}}

\vspace*{-0.3 cm}
\caption{Scenario IV: two bias and ramp attack profiles for the coordinated attack on two different converters; the ramp attack is on the communication.}

\label{fig:coordinated_signal}

\vspace*{-0.3 cm}

\end{figure}

In the last scenario, a coordinated attack is applied on both the sensor measurement and communication lines. The attack signal is depicted in Fig.~\ref{fig:coordinated_signal}. At $t=0.3$~s, a bias attack is launched on the sensor for the fourth converter, and after a while, at $t=0.6$~s, a ramp attack is applied to the communication line from the second source to the third source. The result of a ramp attack on the communication line on the output current with and without mitigation is shown in Fig.~\ref{fig:cordinatedramp5}. $I_{attack-3}$ and $I_{attack-4}$ are the output currents of the converters when the attack is not mitigated. $I_{attack-3}$ and $I_{attack-4}$ are due to ramp attacks on the communication and bias attacks on the sensor, respectively. $I_{o3}$ and $I_{o4}$ are the ones with detection and mitigation. The current fluctuation under normal conditions is 4.68\%. This value increases to 6.79\% after cyberattack mitigation, which is the estimated current using LSTM. 

\begin{figure}[t]
\centerline{\includegraphics[width= 1\columnwidth ]{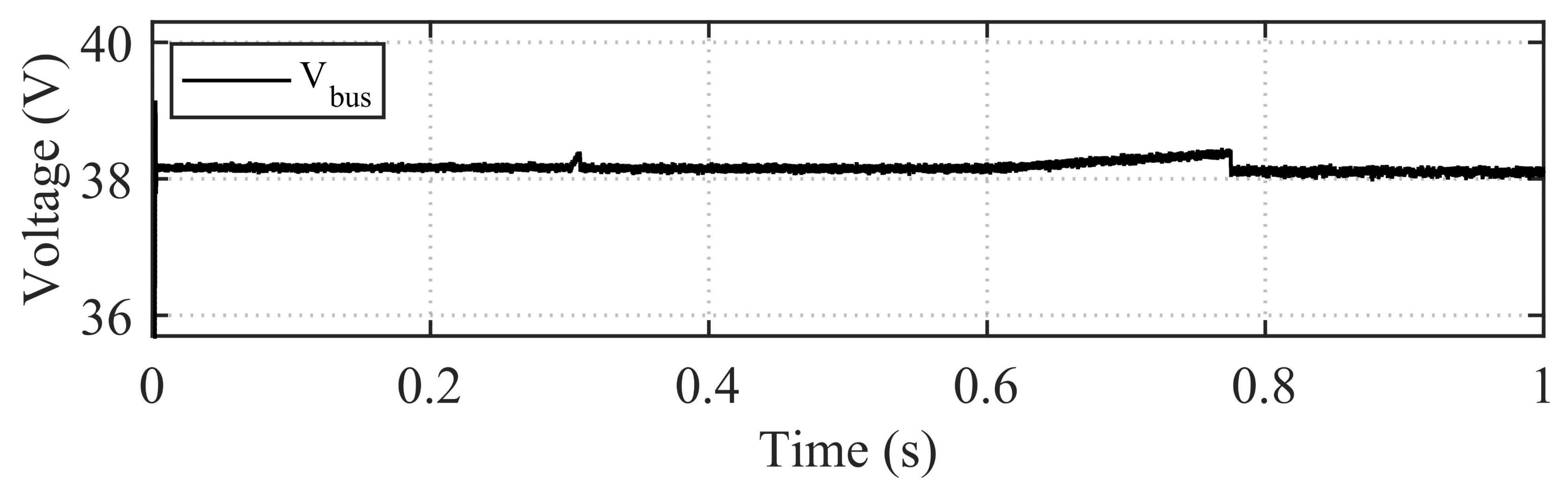}}

\vspace*{-0.2 cm}
\caption{Scenario IV: bus voltage profile after a ramp attack plus bias attack on two converters at $t=0.3$~s and $t=0.6$~s.}
\label{fig:ramp_voltage}

\vspace*{-0.2 cm}
\end{figure}

%Thus, the attack is not mitigated and the current starts to rise until it damages the system or makes the system unstable. 
Fig.~\ref{fig:cordinatedramp5} shows the converter output currents and voltages for this scenario. This figure shows the successful detection and mitigation of the attacks. Additionally, Fig.~\ref{fig:ramp_voltage} shows the bus voltage of the converter for this scenario. This figure shows two anomalies, both of which are mitigated. Therefore, using simulation studies in various scenarios, the proposed hybrid algorithm can successfully detect the FDI attacks with less than 1.4\% error and replace the falsified data with estimated data provided by the LSTM block.

%=======================================================

\section{Conclusion}
% The incorporation of parallel converters into DC microgrids and the requirement for communication between them exposes weaknesses in microgrids.
Manipulation of the neighbor converter command or converter sensor measurement may cause converter instability or damage in a microgrid. Therefore, this paper investigates the intrusion issue, targeting parallel DC-DC converters in a DC microgrid. False data injection attacks on sensors and communication lines in DC microgrids are examined. A hybrid machine learning-based approach is proposed in this paper. The method utilizes logistic regression to detect the anomalies in the system, and after the LSTM cell confirms the detection, it replaces the falsified data with an estimated value for the manipulated signal. Simulation studies are performed within the MATLAB/Simulink environment to validate the effectiveness of the proposed method in various scenarios. The simulation results show the effectiveness of the detection and mitigation of FDIA under various scenarios.

\bibliographystyle{IEEEtran}
\bibliography{ref.bib}

%==================================================
% \printbibliography %Prints bibliography

\end{document}